# Tests of a Novel Design of Resistive Plate Chambers

B. Bilki[a,c], F. Corriveau[d], B. Freund[a,d], C. Neubüser[a,b], Y. Onel[c], J. Repond[a,*], J. Schlereth[a], and L. Xia[a]

[a] Argonne National Laboratory,
   9700 S. Cass Avenue, Argonne, IL 60439, U.S.A.
[b] DESY,
   Notkestrasse 85, D-22607 Hamburg, Germany
[c] University of Iowa,
   Iowa City, IA 52242-1479, U.S.A.
[d] McGill University,
   3600 University Street, Montreal, QC H3A2T8, Canada

E-mail: repond@anl.gov

ABSTRACT: A novel design of Resistive Plate Chambers (RPCs), using only a single resistive plate, is being proposed. Based on this design, two large size prototype chambers were constructed and were tested with cosmic rays and in particle beams. The tests confirmed the viability of this new approach. In addition to showing an improved single-particle response compared to the traditional 2-plate design, the novel chambers also prove to be suitable for calorimetric applications.

---

[*] Corresponding author.

**Contents**

**1. Introduction**

Resistive Plate Chambers (RPCs) were first introduced in the 1980's [1]. Their design typically features two resistive plates made of either Bakelite or glass. The readout board is placed on the outside of the chamber and contains strips or pads which pick up the signals inductively. RPCs are widely used in High Energy Physics experiments, foremost for triggering and precision timing purposes.

In this paper, we propose a novel design based on a single resistive plate, here made of soda-lime float glass. This work was performed in the context of studies of imaging calorimetry for a future lepton collider, as carried out by the CALICE collaboration [2]. The novel design was read out using the standard Digital Hadron Calorimeter (DHCAL) electronic readout system [3] featuring 1 x 1 $cm^2$ signal pads. Tests were performed with both cosmic rays and particle beams.

The novel 1-glass design offers a number of distinct advantages:

- The average pad multiplicity is close to unity, even for particle detection efficiencies approaching 100%. For calorimetric applications, the measured response depends critically on both the efficiency and the average pad multiplicity. With the latter being close to unity the calibration procedure is significantly simplified, compared to what is necessary for a calorimeter utilizing 2-glass chambers as active media.

- The average pad multiplicity usually depends on the value of the surface resistivity of the resistive coating applied to the outside of the anode side of the chamber [4]. But with the corresponding resistive plate eliminated, the strict requirements on the value and uniformity of the resistive coating have become immaterial. The surface resistivity of the coating on the cathode side has only a minor effect on the average pad multiplicity.

- The overall thickness of the device is reduced by the thickness of one of the resistive plates, typically of the order of 1 mm. Thus the thickness of the entire detector including the readout board could be reduced to about 3 mm, with obvious advantages for calorimetry.
- By omitting one resistive plate the rate capability of the chamber is enhanced by approximately a factor two, due to the decrease in the overall bulk resistivity of the chamber.

## 2. Description of the 1-glass RPC design

A sketch of the 1-glass design is shown in Figure 1. The thickness and bulk resistivity of the glass plate was 1.1 mm and $4.7 \times 10^{13}$ $\Omega$cm, respectively. The chambers measure approximately 32 x 48 cm$^2$ and are read out by 1536 readout pads, each with an area of 1 x 1 cm$^2$. The signal pick-up pads are placed directly in the gas volume, are gold plated and separated from each other by small grooves with a width of 220 μm. To guarantee a smooth surface, the vias connecting the pads to the readout system were filled with conductive epoxy and the surface was sanded down to eliminate all dimples and hills.

A front-end ASIC, the DCAL III chip [5], is connected to 64 pads and applies a single threshold to each signal line. The connections to the pads are DC and provide through the input voltage of the pre-amplifier an equipotential surface to well within 1 Volt. The value of the threshold is common to all 64 channels of an ASIC and can be adjusted within the range of approximately 10 to 600 fC. The front-end ASICs are located directly on the readout-board (on the opposite side of the pads) and are read out into an FPGA serving as data concentrator, also located on the same board. The data is collected either in self-triggered mode (for cosmic rays) or subsequent to an external trigger (in the test beam). Including the readout board the chambers have a thickness of 6 mm. Currently, the chambers' thickness is dominated by the readout board. With thinner boards and embedded ASICs the overall thickness could be reduced to about 3 mm.

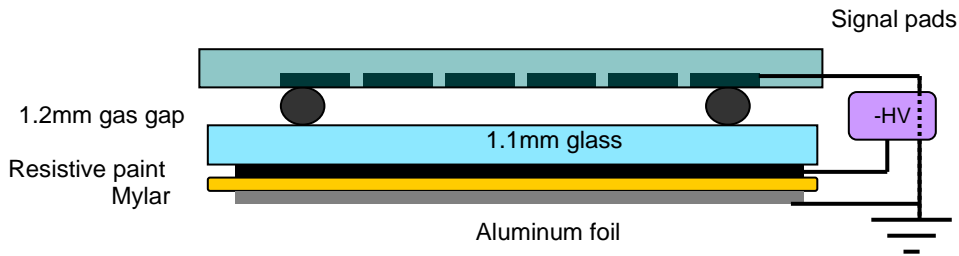

**Figure 1.** Sketch of the design of a 1-glass RPC.

In total two chambers were built based on the 1-glass design. The chambers were operated with a mixture of R134a (94.5%), Isobutane (5.0%) and SF$_6$ (0.5%), which is non-flammable. A high voltage of 7 kV was applied to the resistive coating of the glass plate. Photographs of one of these chambers are shown in Figure 2.

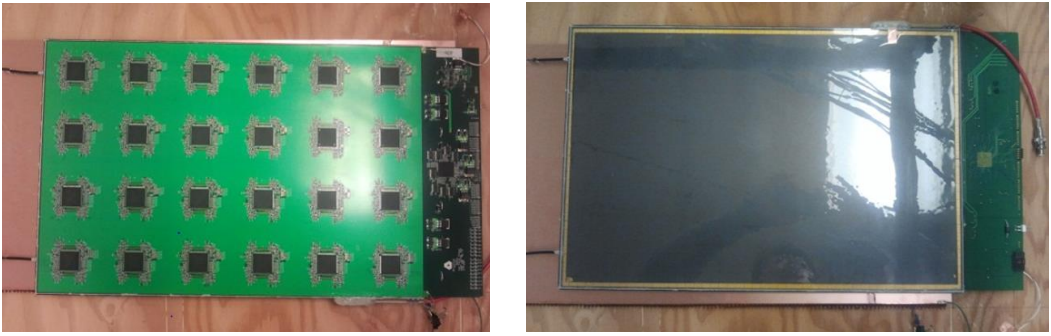

**Figure 2.** Photographs of one of the 1-glass RPCs: (left) the top with the readout ASICs and (right) the cathode plate with the resistive coating.

## 3. Tests with cosmic rays

The two prototype 1-glass RPCs were tested in a cosmic ray test stand. The stand consisted of seven RPCs which were built according to a standard 2-glass design [3], with the two novel RPCs interleaved in the stack. The standard RPCs were used to reconstruct tracks traversing the stack and to determine the position at which they intersect the RPCs under test. The data were collected in trigger-less mode, with a software requirement of hits in at least three of the seven tracking layers within any 700 nanosecond time interval.

The efficiency of the test chambers was determined by searching for clusters of hits within 2 cm of the interpolated position of the cosmic ray track, as reconstructed by the other chambers in the stack. (The clustering was performed by a nearest neighbor clustering algorithms, which required a common edge between 2 hits in order to belong to the same cluster.) If such a cluster was found, the pad multiplicity was defined as the number of hits belonging to the cluster. For one of the chambers and as function of lateral position, Figures 3 and 4 show maps of the efficiency and the average pad multiplicity, respectively. The chamber exhibits a uniform efficiency with an average of 95%. As expected the average pad multiplicity is also uniform and close to unity. The left bottom corner exhibits a higher pad multiplicity due to accidental noise hits occurring in the vicinity of the ground connection of the chamber. Results obtained with the other prototype chamber were very similar.

## 4. Noise rates

The noise rate in the one-glass RPCs was estimated by collecting self-triggered data over intervals of 60 seconds each. At the default operating conditions with a high voltage of 7 kV the accidental hit rate was estimated to be $0.2 - 0.3$ Hz/cm$^2$. This rate is comparable or slightly higher to the corresponding rate in RPCs based on the standard 2-glass design.

## 5. Tests in a particle beam

The two prototype chambers were placed into the 120 GeV primary proton beam of the Fermilab Test Beam Facility [6]. At the facility, the test beam arrives in spills of approximately 3.5 second duration. The beam was slightly defocused, such that the area of incidence onto the chambers was easier to establish. A wire chamber placed upstream of the chambers measured the beam spot with a standard deviation of 2 – 5 mm in both horizontal and vertical directions. The beam intensity was measured with a set of five scintillator counters, also placed upstream of the test chambers. For a given beam setting, the rates measured by the different counters proved to be internally consistent. Data was collected in separate runs, each with a specific beam intensity. The latter was varied between 500 and 1,000,000 counts per spill. Assuming an overall beam spot of two standard deviations in each direction, these beam intensities in turn correspond to particle rates of 300 Hz/cm$^2$ to 500 kHz/cm$^2$ impinging onto the prototype chambers. A systematic uncertainty of a factor of 2 was assigned to the rate estimate, due a) to the uncertainty in defining the size of the area of particle incidence, b) to variations of the beam intensity of up to a factor of two within a given spill, and c) to uncertainties in relating the measured effects to the case of uniform illumination of the chambers.

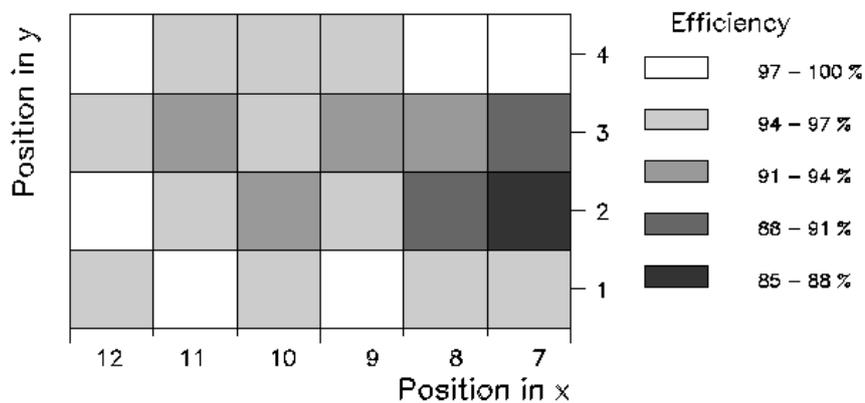

**Figure 3.** Detection efficiency for cosmic ray particles as function of position on the 1-glass RPC.

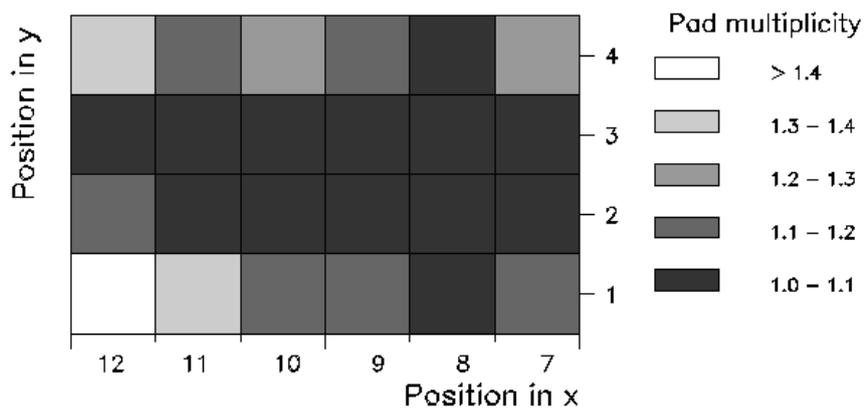

**Figure 4**. Average pad multiplicity measured for cosmic ray tracks as function of position on the 1-glass RPC.

The readout of the chambers was triggered by the coincidence of two scintillator counters placed upstream of the chambers. The particle detection efficiency and average multiplicity were calculated in a similar way as for cosmic rays. The efficiency was calculated as the ratio of the number of events with at least one hit in a given chamber and the total number of triggers. Figure 4 shows the results as function of particle flux. As can be seen from the figure both chambers performed in a very similar way. The efficiency measured at the lowest rate achieved in the test beam, corresponding to 340 Hz/cm$^2$, is observed to be only slightly lower than the efficiency of 95% measured with cosmic rays. For rates above 300 Hz/cm$^2$ the efficiency decreases to reach a minimum below 20% at the highest beam intensities probed. For rates above 5 kHz/cm$^2$ the average pad multiplicity is seen to increase, probably due to the fact that particles which initiated a hadronic shower upstream of the chambers create higher particle multiplicities. These in turn generate additional avalanches in the gas gap which will increase the signal charge induced in the readout pads, thus increasing the probability of registering a hit.

In order to validate the chambers as candidates for the active medium of an imaging hadronic calorimeter, a block of tungsten with a thickness corresponding to one nuclear interaction length was placed upstream of the chambers. Figure 5 shows the number of hits observed in one chamber versus the number of hits seen in the other. As expected, a clear correlation is seen, corresponding to a Pearson correlation factor of 0.90.

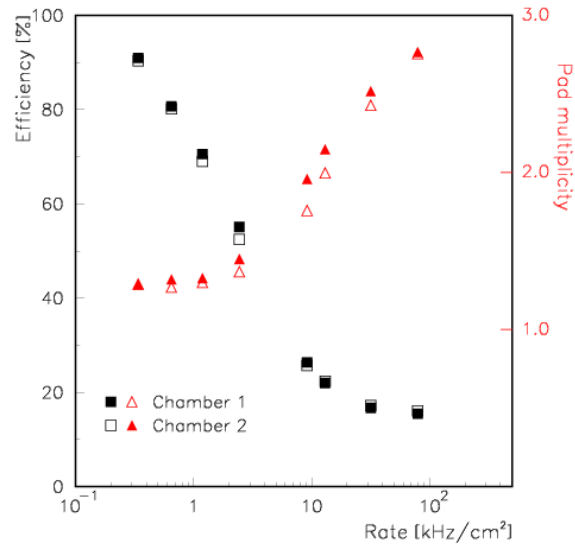

**Figure 4.** MIP detection efficiency (black) and average pad multiplicity (red) as measured versus particle rate with the 120 GeV primary proton beam at the Fermilab Test Beam Facility.

**6. Conclusions**

Resistive Plate Chambers, based on a novel design with a single resistive (glass)-plate, have been tested with both cosmic rays and particle beams. The chambers were read out with 1 × 1 cm$^2$ pads. The particle detection efficiency of the chambers was measured to be approximately 95% for average pad multiplicities close to unity. The rate capability of the chambers was established in a 120 GeV proton beam and showed the expected drop off of efficiency for particle rates above 300 Hz/cm$^2$. When placed in the particle beam, but behind a block of Tungsten, the two prototype chambers show a strong correlation in the number of hits, validating their suitability as the active medium of a hadron calorimeter.

The novel chamber design offers a number of advantages over the traditional two-plate design: an average pad multiplicity close to unity, a reduced overall thickness, and a simplified construction procedure. The latter is mostly due to the relaxed requirement for a specific surface resistivity of the resistive layer and the omission of one glass plate.

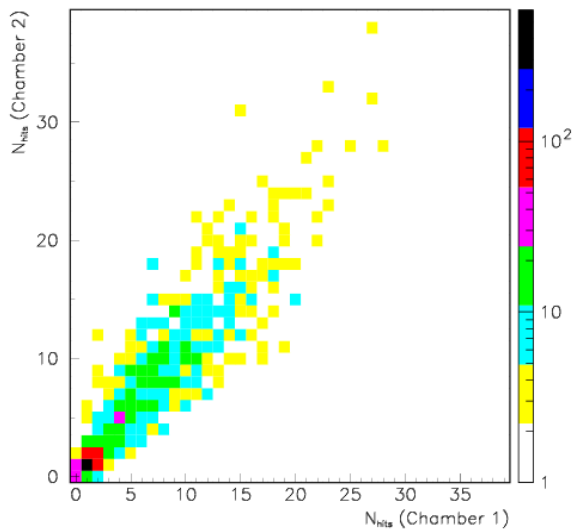

**Figure 5.** Number of hits in chamber 2 versus the number of hits in chamber 1. The chambers were placed behind one interaction length of Tungsten and were exposed to the 120 GeV primary proton beam at the Fermilab Test Beam Facility.

## References

[1]  R. Santonico and R. Cardarelli, Nucl. Instr. and Meth. 187 (1981)377-380

[2]  https://twiki.cern.ch/twiki/bin/view/CALICE/WebHome

[3]  Design, Construction, and Testing of the Digital Hadron Calorimeter (DHCAL) Electronics (paper in preparation)

[4]  G. Drake et al., Nucl. Instr. and Meth. A578 (2007) 88

[5]  A. Bambaugh et al., *Production and commissioning of a large prototype Digital Hadron Calorimeter for future colliding beam experiments,* Nuclear Science Symposium and Medical Imaging Conference (NSS/MIC), 2011 IEEE; DOI: 10.1109/NSSMIC.2011.6154437

[6]  http://www-ppd.fnal.gov/FTBF/